\documentclass[a4paper,10pt,twoside]{article}
\usepackage{amsmath,amsfonts,amssymb,amsthm,multirow}
\usepackage{graphicx}
\usepackage{grffile}
\usepackage{epstopdf,url}
\usepackage {epsfig}

\newtheorem{theorem}{Theorem}

\title{Estimating an incidence rate based on two imperfect observers}
\author{Tamar Gadrich{\footnotemark[1]},\; Guy Katriel{\footnotemark[2]}}

\date{}

\begin{document}

\title{Estimating the rate of defects under imperfect sampling inspection - a new approach%\thanks{Grants or other notes
}


%\author{Tamar Gadrich{\footnotemark[1]},\; Guy Katriel{\footnotemark[2]}}


\maketitle
\renewcommand{\thefootnote}{\fnsymbol{footnote}}
\footnotetext[1]{ Department of Industrial Engineering and Management, ORT Braude College, Karmiel, Israel.}
\footnotetext[2]{ Department of Mathematics, ORT Braude College, Karmiel, Israel.}

\begin{abstract}
	We consider the problem of estimating the rate of defects (mean
	number of defects per item), given the counts of defects detected by two independent imperfect inspectors on one sample of items. In contrast with the setting for the well-known method
	of Capture-Recapture, we {\it{do not}} have information 
	regarding the number of defects jointly detected by {\it{both}} inspectors. 
	We solve this problem by constructing two types of estimators - a simple moment-type estimator, and a complicated maximum-likelihood estimator.
	The performance of these estimators is studied analytically and by means of simulations. It is shown that the 
	maximum-likelihood estimator is superior to the moment-type estimator. A systematic comparison
	with the Capture-Recapture method is also made.
	
%	\keywords{estimation \and incidence \and partial reporting \and moment estimators \and maximum likelihood }
	
\end{abstract}

\section{Introduction}

Given a sample of noncomforming items, each of which may contain several defects, we would like to estimate the defect rate $\lambda$ -
the mean number of defects per item, in the population from which the items are sampled. 
Such problems are at the heart of statistical quality control (Montgomery \cite{montgomery}), where an estimate of the defect rate 
is used to accept or reject the lot that the sample was taken from. With perfect inspection, 
basic statistical theory allows us to estimate $\lambda$ and give an appropriate confidence interval, depending on the
number of items inspected.

When inspection is {\it{imperfect}}, {\it{i.e.}} when some defects may not be identified, things become more difficult.
If it is assumed that the detection rate $p\in (0,1)$ (the probability that each defect is detected) is known, then one may adjust the estimate of
the defect rate and corresponding confidence interval accordingly (Kotz and Johnson \cite{kotz}). However if the rate of detection is unknown, one cannot construct a consistent estimator for the defect rate based
on the counts the number of defects detected in each item - in statistical terminology, the problem of
joint estimation of $\lambda$ and of $p$ is {\it{unidentifiable}}. 

A way around this difficulty is using the idea of `Capture-Recapture', originated from the field of ecology for estimating animal
populations (Chao \cite{chao2}, McCrea and Morgan \cite{mcrea}). The idea is that an item is inspected twice (or more times), independently. 
For each item, the number of defects found by each of the inspectors, as well as the number of joint detections - defects found by {\it{both}} 
inspectors, is recorded. Using independence of the two inspections, it is possible to estimate both the true number of defects {\it{and}} the detection rates 
of the two inspectors.  This approach has been widely used in the field of software testing (Briand et al. \cite{briand}). A related but different
approach is that of serial inspection, in which it is assumed that defects detected by one inspector are removed 
before an item is inspected by a second inspector (Bonett and Woodward \cite{bonnet1}, Chun \cite{chun}).

In this work we deal with an even more challenging situation, in which items are inspected by two inspectors (human or electronic), but the
available information consists only of the counts of the number of defects detected by each of the inspectors in each of the items -
we assume that the number of {\it{joint detections}}  is unknown. 
Such a situation occurs when it is difficult to record the position of a defect on the item in such a way that
it will be possible to match the defects detected by the two inspectors. For example, if an item consists of 
a large number of identical subunits, which may not remain in a fixed position relative to one another, it may be impossible to determine the defective subunits detected by both inspectors.
If automated optical inspection is performed, with each item scanned by two cameras, it may be difficult
for a pattern recognition system to determine the number of joint detections.

We show here that, even in the absence of data pertaining to joint detections, it is still possible to 
estimate the defect rate $\lambda$, together with the detection rates $p_1,p_2$ of the two inspectors.

We have found no previous mention of the possibility to jointly estimate defect and detection rates without 
data on joint detections in the literature concerning counting with imperfect
detection. There is, however, an analogous problem concerning measurements of continuous characteristics, {\it{e.g.}} weights of 
	objects, which has received some attention. Given a set of objects sampled from a population of objects whose weights are normally
	distributed with unknown mean and variance, assume the weight of each object is measured by two (or more) different measuring 
	instruments with normally-distributed measurement errors that have means $0$ and unknown variances. Grubbs \cite{grubbs} (see also 
	\cite{grubbs1,lombard}) has shown that such data is sufficient in order to estimate both the population mean and variance and the measurement error variances (whereas a single measurement
	of each object would obviously not suffice for separating population variance from measurement error variance). The method
	proposed here can be seen as the application of an analogous approach to discrete data.

We will develop two types of estimators for the defect rate $\lambda$ and the detection rates $p_1,p_2$: a simple moment-type estimator and a more
complicated maximum likelihood estimator. 
The moment-type estimators, to be presented in Section \ref{moment},
lead to simple explicit formulas, immediately showing that the parameters are indeed identifiable.
These estimators are amenable to detailed analysis, which will provide asymptotic expressions for 
the bias and variance of the estimators, and asymptotic confidence intervals. These allow one to determine the number
of items which need to be inspected in order to achieve the desired precision in the estimate of defect rate.

The method of maximum-likelihood estimation, developed in Section \ref{mlest},
leads to a relatively complicated maximization problem which cannot be solved in closed form. We will however show that the three-variable maximization problem can be reduced to a one-variable equation, so that numerical solution
of the problem is straightforward.

We will compare the two methods by numerical simulation, applying both to simulated data generated from our model, and show that the maximum likelihood estimator is (on the average)  more precise than the moment estimator. Thus the higher complexity of estimation using the maximum likelihood method is re-payed by better estimates. 

We will also compare our method with the aforementioned Capture-Recapture method. As our method requires less
information than the Capture-Recapture method, it is to be expected that it will require more items to be inspected in order to 
achieve an estimate at the same level of precision. We will examine this quanititatively in Section \ref{comparison}.

\section{The Model}
\label{model}

Here we define the statistical model and formulate the estimation problem involved.

We assume that the number of defects in an item is Poisson-distributed with mean $\lambda$. 
Our main task is to estimate the parameter $\lambda$, given a sample of $M$ items ($M\geq 2$).

We denote by $N_m$ the (unknown) number of defects in item $m$ ($1\leq m\leq M$). Thus, by our assumption,
$N_m$ are independent random variables with the distribution
\begin{equation}\label{poi1}N_m\sim Poisson(\lambda),\;\;\;1\leq m\leq M.\end{equation}
Two independent inspectors examine each item, with inspector $i$ ($i=1,2$)  detecting each of the defects with probability $p_i$, independently of the other defects.
Thus the number of defects detected by the inspectors in item $m$ are distributed according to
\begin{equation}\label{poi2}R_{m,i}|N_{m}\sim Binomial (N_m,p_i),\;\;i=1,2,\end{equation}
with $R_{m,1},R_{m,2}$ independent {\it{conditional}} on $N_m$.

We assume that the detection rates $p_1,p_2$, as well as the defect rate $\lambda$, are {\it{not known}} to us. 

Our problem is:
\begin{center}
	Given that 
	$R_{m,i}=r_{m,i}$, $(m=1,\cdots, M, \;i=1,2)$,
	estimate the defect rate $\lambda$ and the detection rates $p_1,p_2$.
\end{center}

As was mentioned in the Introduction, if we had only one inspector who detects a defect with probability $p$, then
the numbers of defects detected on the $M$ items would be a sample from a Poisson distribution with mean $\lambda p$, and there would have been no
way to use the observation data to separately estimate $\lambda$ or $p$ - the parameters are non-identifiable.
The key observation here is that given the counts of defects detected by two independent imperfect inspectors, the parameters 
$\lambda,p_1,p_2$ become identifiable. This will be shown in the next section by constructing simple explicit estimators which are consistent,
that is, converge in probability to the true values of the parameters as the number of inspected items $M$ tends to infinity.

We introduce some further notation that will be useful. Denote:
\begin{itemize}
	\item $X_{m,i}$ is the number of defects detected {\it{only}} by inspector $i$ in item $m$.
	
	\item $Y_m$ is the number of defects detected by {\it{both}} inspectors in item $m$.
\end{itemize}

Note that
\begin{equation}\label{ss}
R_{m,i}=X_{m,i}+Y_m,\;\;\;i=1,2.
\end{equation}
Using Poission decomposition it follows that $\{X_{m,1},X_{m,2},Y_m\}_{m=1}^M$ are {\it{independent}} and Poisson distributed with
\begin{equation}\label{x0}X_{m,1}\sim Poisson(\lambda p_1(1-p_2)),\;\;\;X_{m,2}\sim Poisson(\lambda p_2(1-p_1)),\end{equation}
\begin{equation}\label{y0}Y_{m}\sim Poisson(\lambda p_1p_2).\end{equation}
To see this, note that each defect is detected only by inspector $1$ with probability $p_1(1-p_2)$, only by inspector $2$ with probability $p_2(1-p_1)$, by
both inspectors with probability $p_1p_2$, and by none of them with probability $(1-p_1)(1-p_2)$. 

Our problem can thus be re-stated as: 

\begin{center}
	Given the values $\{ r_{m,i}\}_{m=1}^M$ ($i=1,2$) of the random variables, $\{ R_{m,i}\}_{m=1}^M$, 
	generated by (\ref{ss}),(\ref{x0}),(\ref{y0}), with the independence assumption on the set $\{X_{m,1},X_{m,2},Y_m\}_{m=1}^M$,
	estimate $\lambda,p_1,p_2$.
\end{center}

Note that $R_{m,1},R_{m,2}$ are {\it{dependent}} (although they are {\it{conditionally independent}} given the value $N_m$),
but the sequence  $\{(R_{m,1},R_{m,2})\}_{m=1}^M$ is an independent and identically distributed sequence of two-dimensional 
random variables. We can therefore, when convenient, remove the subscript $m$ and refer to random variables $R_1,R_2$, which are given by
$$R_1=X_1+Y,\;\;\;R_2=X_2+Y,$$
where $(X_1,X_2,Y)$ are independent, with
\begin{equation}\label{x}X_{1}\sim Poisson(\lambda p_1(1-p_2)),\;\;\;X_{2}\sim Poisson(\lambda p_2(1-p_1)),\end{equation}
\begin{equation}\label{y}Y\sim Poisson(\lambda p_1p_2).\end{equation}
Our data $\{(r_{m,1},r_{m,2})\}_{m=1}^M$ can thus be considered as $M$ independent realizations of the vectorial random variable $(R_1,R_2)$.

It should be noted that the distribution of the pair $(R_1,R_2)$ is known as the {\it{Bivariate Poisson}}
distribution with parameters $\theta_1=\lambda p_1(1-p_2),\theta_2=\lambda p_2(1-p_1),\theta_{12}= \lambda p_1p_2$ (Johnson et al. \cite{johnson}).
Estimation of the parameters $\theta_1,\theta_2,\theta_{12}$ for a Bivariate Poisson random variable, based on a 
random sample, has been considered by Holgate \cite{holgate} (see also Kocherlakota \& Kocherlakota \cite{kocherlakota}), who obtained both moment and maximum-likelihood type estimators.
In fact, our estimators for $p_1,p_2,\lambda$ can be directly related to those of Holgate for $\theta_1,\theta_2,\theta_{12}$,
through the relation between the sets of variables. However, the results of the {\it{analysis}} of the moment estimators for $\lambda,p_1,p_2$, {\it{i.e.}} 
the approximate bias and variance (see section \ref{analysis} below) do {\it{not}} follow from the corresponding analysis made in Holgate \cite{holgate}
for the estimators of the Bivariate Poisson parameters, because the nonlinear relationship between the two parametrizations entails 
that computing asymptotic formulas for the moments of first and second order of the estimators of the parameters $\lambda,p_1,p_2$ requires 
computation of moments of {\it{higher order}} for the estimators of $\theta_1,\theta_2,\theta_{12}$ (see section \ref{analysis} below).

\section{Moment-type estimators}
\label{moment}

\subsection{Obtaining the estimators}

We have, using (\ref{ss})-(\ref{y0}),
$$R_{i}\sim Poisson(\lambda p_i),\;\;\; i=1,2.$$
so that
$$E(R_{1})= \lambda p_1,\;\;\;E(R_{2})= \lambda p_2.$$
Also, using independence, and by (\ref{y}),
$$COV(R_{1},R_{2})=COV\left( X_{1}+Y,X_{2}+Y\right)= VAR(Y)=\lambda p_1 p_2,$$
so that
$$p_1=\frac{COV(R_{1},R_{2})}{E(R_{2})},\;\;\; p_2=\frac{COV(R_{1},R_{2})}{E(R_{1})},\;\;\;\lambda= \frac{E(R_{1})E(R_{2})}{COV(R_{1},R_{2})}.$$

We can therefore estimate $E(R_{1}),E(R_{2}),COV(R_{1},R_{2})$ by the corresponding unbiased empirical mean and co-variance estimators
$$E(R_{i})\approx \bar{r}_i\doteq\frac{1}{M}\sum_{m=1}^M r_{m,i},\;\;\;i=1,2$$
$$COV(R_{1},R_{2})\approx \hat{S}_{1,2}\doteq\frac{1}{M-1}\sum_{m=1}^M (r_{m,1}-\bar{r}_1)(r_{m,2}-\bar{r}_2),$$
and obtain estimators for $p_1,p_2,\lambda$ given by
\begin{eqnarray}\label{estimators}\hat{p}_1&=&\frac{\hat{S}_{1,2}}{\bar{r}_2}=\frac{\frac{1}{M-1}\sum_{m=1}^M (r_{m,1}-\bar{r}_1)(r_{m,2}-\bar{r}_2)}
{\frac{1}{M}\sum_{m=1}^M r_{m,2}},\nonumber\\
\hat{p}_2&=&\frac{\hat{S}_{1,2}}{\bar{r}_1}=\frac{\frac{1}{M-1}\sum_{m=1}^M (r_{m,1}-\bar{r}_1)(r_{m,2}-\bar{r}_2)}{\frac{1}{M}\sum_{m=1}^M r_{m,1}},\nonumber\\
\hat{\lambda}&=&\frac{\bar{r}_1\cdot \bar{r}_2}{\hat{S}_{1,2}} =\frac{\left(\frac{1}{M}\sum_{m=1}^M r_{m,1} \right)\cdot  \left(\frac{1}{M}\sum_{m=1}^M r_{m,2} \right)}{\frac{1}{M-1}\sum_{m=1}^M (r_{m,1}-\bar{r}_1)(r_{m,2}-\bar{r}_2)}.\end{eqnarray}
%an event with probability
%$$P(r_{1,i}=r_{2,i}=\cdots = r_{M,i}=0)=(P(R_{i}=0))^M=e^{-\lambda p_i M}\xrightarrow{M\rightarrow\infty} 0,$$
%and the estimator $\hat{\lambda}$ is well-defined except in the case that all $r_{1,i}=\cdots=r_{t,i}$ for either $i=1$ or $i=2$,
%an event which has probability
%$$P(r_{1,i}=r_{2,i}=\cdots = r_{M,i})=\sum_{k=0}^\infty (P(R_{i}=k))^M=e^{-\lambda p_i M}\sum_{k=0}^\infty \left(\frac{\lambda^k}{k!} \right)^M,\;\;\;$$
%$$=e^{-\lambda p_i M}\left[\sum_{\frac{(2\lambda)^k}{k!}\geq 1} \left(\frac{\lambda^k}{k!} \right)^M +\sum_{\frac{(2\lambda)^k}{k!}< 1} \left(\frac{\lambda^k}{k!} \right)^M \right]$$
%$$\leq e^{-\lambda p_i M}\left[\sum_{\frac{(2\lambda)^k}{k!}\geq 1} \left(\frac{\lambda^k}{k!} \right)^M +\sum_{\frac{(2\lambda)^k}{k!}< 1} \left(\frac{1}{2^M} \right)^k\right] $$
%$$\leq e^{-\lambda p_i M}\left[\left(\max_{(2\lambda)^k\geq k!} \frac{\lambda^k}{k!} \right)^M +1\right]
%\xrightarrow{M\rightarrow \infty} 0.$$
%Mhus the probability that the estimators are defined rapidly approaches $1$ as $M$ increases.

By the weak law of large numbers ({\it{e.g.}} Mood et al. \cite{mood}) we have convergence in probability, as $M\rightarrow \infty$:
$$\bar{r}_i \xrightarrow{P} E(R_i)=\lambda p_i,\;\;\;\;i=1,2,\;\;\;\hat{S}_{1,2}  \xrightarrow{P} COV(R_{1},R_{2})=\lambda p_1 p_2,$$
implying 
$$\hat{p}_i\xrightarrow{P}p_i\;\;\; i=1,2,\;\;\;\;\hat{\lambda}\xrightarrow{P} \lambda.$$
Thus the estimators $\hat{p}_1,\hat{p}_2,\hat{\lambda}$ are consistent, which in particular shows that the parameters 
$p_1,p_2,\lambda$ of our problem are identifiable.

Let us note that the estimator $\hat{p}_1$ is undefined when $\bar{r}_2=0$, the estimator $\hat{p}_2$ is undefined when $\bar{r}_1=0$, 
and the estimator $\hat{\lambda}$ to be undefined when $\hat{S}_{1,2}=0$. However since both $\bar{r}_i$ and $\hat{S}_{1,2}$
are asymptotically normal with nonzero means ($E(R_i)$ and $COV(R_1,R_2)$, respectively) and variances of order $O(\frac{1}{M})$
(for $\bar{r}_i$ this is immediate, for $\hat{S}_{1,2}$ see (\ref{holgate}) below), these events
have  probability which is exponentially small in $M$ as $M\rightarrow \infty$, so that they do not have practical impact.

\subsection{Asymptotic bias and variance of the moment estimators}
\label{analysis}

In the following, we obtain the asymptotic expectation and variance of the moment estimators, in order to assess their bias and 
their precision, and to derive confidence intervals.  We later check by simulation that these
asymptotic approximations are satisfactory even for moderate values of items $M$.

The calculations are based on the approximations for the mean of a ratio of two random variables. From (\ref{estimators}), using second-order Taylor expansions (see, {\it{e.g.}}, Mood et al. \cite{mood}), these are given by:
\begin{eqnarray*}
E(\hat{p}_1) &=& \frac{E(\hat{S}_{1,2})}{E(\bar{r}_2)} - \frac{COV(\hat{S}_{1,2},\bar{r}_2)}{E^2(\bar{r}_2)} +\frac{VAR(\bar{r}_2) \cdot E(\hat{S}_{1,2})}{E^3(\bar{r}_2)}+ O\left(\frac{1}{M^2}\right), \\
E(\hat{p}_2) &=& \frac{E(\hat{S}_{1,2})}{E(\bar{r}_1)} - \frac{COV(\hat{S}_{1,2},\bar{r}_1)}{E^2(\bar{r}_1)} +\frac{VAR(\bar{r}_1) \cdot E(\hat{S}_{1,2})}{E^3(\bar{r}_1)}+ O\left(\frac{1}{M^2}\right), \\
E(\hat{\lambda})&=& \frac{E(\bar{r}_1 \cdot \bar{r}_2)}{E(\hat{S}_{1,2})}-\frac{COV(\bar{r}_1 \cdot \bar{r}_2,\hat{S}_{1,2})}{E^2(\hat{S}_{1,2})}+\frac{VAR(\hat{S}_{1,2}) \cdot E(\bar{r}_1 \cdot \bar{r}_2)}{E^3(\hat{S}_{1,2})}+ O\left(\frac{1}{M^2}\right),
\end{eqnarray*}
\begin{eqnarray*} 
VAR(\hat{p}_1) &=& \frac{E^2(\hat{S}_{1,2})}{E^2(\bar{r}_2)} \left[\frac{VAR(\hat{S}_{1,2})}{E^2(\hat{S}_{1,2})}-2 \frac{COV(\hat{S}_{1,2},\bar{r}_2)}{E(\hat{S}_{1,2}) \cdot E(\bar{r}_2)} +\frac{VAR(\bar{r}_2)}{E^2(\bar{r}_2)}\right]+O\left(\frac{1}{M^2}\right),\\
VAR(\hat{p}_2) &=& \frac{E^2(\hat{S}_{1,2})}{E^2(\bar{r}_1)} \left[\frac{VAR(\hat{S}_{1,2})}{E^2(\hat{S}_{1,2})}-2 \frac{COV(\hat{S}_{1,2},\bar{r}_1)}{E(\hat{S}_{1,2}) \cdot E(\bar{r}_1)} +\frac{VAR(\bar{r}_1)}{E^2(\bar{r}_1)}\right]+O\left(\frac{1}{M^2}\right), \\
VAR(\hat{\lambda})&=& \frac{E^2(\bar{r}_1 \cdot \bar{r}_2)}{E^2(\hat{S}_{1,2})}\left[\frac{VAR(\bar{r}_1 \cdot \bar{r}_2)}{E^2(\bar{r}_1 \cdot \bar{r}_2)}-2 \cdot \frac{COV(\bar{r}_1 \cdot \bar{r}_2,\hat{S}_{1,2})}{E(\bar{r}_1 \cdot \bar{r}_2) \cdot E(\hat{S}_{1,2})}+\frac{VAR(\hat{S}_{1,2})}{E^2(\hat{S}_{1,2})}\right]+O\left(\frac{1}{M^2}\right).\end{eqnarray*}

Some of the terms in the expressions above have already been computed in previous work (Holgate \cite{holgate}, Kocherlakota \& Kochelakota \cite{kocherlakota}), using different notation, namely
\begin{eqnarray}\label{holgate}VAR(\hat{S}_{1,2})&=&\frac{\lambda p_1p_2[\lambda(p_1p_2+1)+1]}{M}+O\left(\frac{1}{M^2}\right),\\
COV(\bar{r}_1,\hat{S}_{1,2})&=&COV(\bar{r}_2,\hat{S}_{1,2})=\frac{\lambda p_1 p_2}{M},\nonumber\end{eqnarray}
though for completeness we give full derivations of these results in our Supporting Material (Gadrich and Katriel \cite{gadrich}, Section 2).
However, two of the terms in the above expressions cannot be obtained from previous work and are obtained in \cite{gadrich}, Section 
3:
\begin{eqnarray*}COV(\bar{r}_1\cdot\bar{r}_2,\hat{S}_{1,2})&=&\frac{1}{M}p_1 p_2\lambda^2 \left(p_1+p_2\right)+O\left(\frac{1}{M^2}\right),\\
VAR(\bar{r}_1\cdot\bar{r}_2)&=&\frac{\lambda^3 p_1p_2}{M}[p_1+p_2+2p_1p_2]+O\left(\frac{1}{M^2}\right).\nonumber
\end{eqnarray*}
The computations involved are surprisingly arduous, but the final results, given below, are 
quite simple. 
\begin{theorem}\label{exp}
	Expectations of the moment estimators are given, as $M\rightarrow \infty$, by
	$$E(\hat{p}_1) =p_1+ O\left(\frac{1}{M^2}\right),\;\;E(\hat{p}_2) =p_2+ O\left(\frac{1}{M^2}\right),$$
	$$E(\hat{\lambda})=\lambda
	+ \left((\lambda+1)\left(1+\frac{1 }{p_1 p_2}\right)-\frac{1}{p_1}-\frac{1}{p_2}\right)\cdot \frac{1}{M}+O\left(\frac{1}{M^2}\right).$$
\end{theorem}
\begin{theorem}\label{variance}
	The variances of the moment estimators are given, as $M\rightarrow\infty$, by
	$$VAR(\hat{p}_i) =\frac{p_i^2}{p_1p_2}\left[1+p_1p_2 +\frac{1
		- p_i}{\lambda}  \right]\cdot \frac{1}{M} + O\left(\frac{1}{M^2}\right),\;\;i=1,2,$$
	$$VAR(\hat{\lambda})=\lambda \left[\lambda\cdot \left(\frac{1}{p_1p_2}+1\right)+ \left(\frac{1}{p_1}-1 \right)\left(\frac{1}{p_2}-1\right)+1 \right]\cdot \frac{1}{M}+O\left(\frac{1}{M^2}\right).$$
\end{theorem}

From these results we see that the precision of the estimator degrades when $p_1$ or $p_2$ is close to $0$.

In Table 1 we compare the approximations obtained in Theorems \ref{exp},\ref{variance} with results obtained 
by simulation: $10^5$ data samples were generated in each simulation, with parameters $\lambda=10$, $p_1=0.4$, $p_2=0.7$, and 
the estimator $\hat{\lambda}$ was computed for each. The good agreement of the simulation results and the approximations is evident,
as well as the fact that as the number of items $M$ increases, the agreement is improved, and the bias of the estimator decreases.

\begin{table}\label{tab1}
	\begin{center}
		\begin{tabular}{|c|c|c|c|c|}
			%	\hline {} & {$\hat{\lambda}$} & {$\hat{p}_1$} & {$\hat{p}_2$} \\ 
			\hline {M} & {$E(\hat{\lambda})$ (sim.)} & {$E(\hat{\lambda})$ (app.)} & {$Std(\hat{\lambda})$ (sim.)} & {$Std(\hat{\lambda})$ (app.)} \\ 
			\hline	100 & 10.51 & 10.46 & 2.54 & 2.18\\
			\hline	200 & 10.25 & 10.23 & 1.65 & 1.54 \\
			\hline	500 & 10.09 & 10.09 & 1.00 & 0.97\\
			\hline 
		\end{tabular}
	\end{center} 
	\caption{Expectation and standard deviation of the moment estimator $\hat{\lambda}$ based on $10^5$ simulations, and the 
		corresponding approximations as given by Theorems \ref{exp},\ref{variance}, for parameters $\lambda=10$, $p_1=0.4$, $p_2=0.7$.}
\end{table}

Using the above results we can give approximate confidence intervals for the parameters at confidence  level $\alpha$:
Let $z^*$ be given by $\Phi^{-1}(1-\frac{\alpha}{2})=z^*$ where $\Phi$ is the cumulative distribution function of the standard normal distribution.
Then confidence intervals for $p_1,p_2$ are given by
$$CI_{p_i}=\left[\hat{p}_i-z^* \cdot SE_{\hat{p}_i},\hat{p}_i+z^* \cdot SE_{\hat{p}_i}\right],\;\;\;i=1,2,$$
where
$$SE_{\hat{p}_i}=\hat{p}_i\sqrt{\frac{1}{\hat{p}_1 \hat{p}_2}\left[1+\hat{p}_1\hat{p}_2 +\frac{1
		- \hat{p}_i}{\hat{\lambda}}  \right]\cdot \frac{1}{M}},\;\;i=1,2,$$
and a confidence interval for $\lambda$ is given by
$$CI_{\lambda}=\left[\hat{\lambda}-z^* \cdot SE_{\hat{\lambda}},\hat{\lambda}+z^* \cdot SE_{\hat{\lambda}}\right],$$
where
$$SE_{\hat{\lambda}}= \sqrt{\hat{\lambda} \left[\hat{\lambda}\cdot \left(\frac{1}{\hat{p}_1\hat{p}_2}+1\right)+ \left(\frac{1}{\hat{p}_1}-1 \right)\left(\frac{1}{\hat{p}_2}-1\right)+1 \right]\cdot \frac{1}{M}}.$$

\section{Maximum likelihood estimation}
\label{mlest}

\subsection{Derivation of maximum likelihood estimators}

In this section we develop an alternative approach to our estimation problem, based on the maximum likelihood method.

Given the data $r_{m,i}$ ($m=1,\cdots,M$, $i=1,2$), and using the shorthand
$$P(r_1,r_2)\doteq P(R_1=r_1,R_2=r_2),$$
(note that $P(r_1,r_2)$ in fact depends also on $\lambda,p_1,p_2$), 
the log-likelihood function (dividing by $M$) is given by
$$LL(\lambda,p_1,p_2)=\frac{1}{M}\ln P(R_{m,i}=r_{m,i},\;1\leq m\leq M,\;i=1,2)=\frac{1}{M}\sum_{m=1}^M \ln P(r_{m,1},r_{m,2}).$$
Using (\ref{ss}), the independence of $X_{1},X_{2},Y$, and (\ref{x}),(\ref{y}), we get
$$P(r_1,r_2)=P(X_{1}+Y=r_{1},X_{2}+Y=r_{2})$$
$$= \sum_{l=0}^{\min(r_{1},r_{2})} P(Y=l)P(X_{1}=r_{1}-l)P(X_{2}=r_{2}-l).$$
$$=e^{-\lambda[1-(1-p_1)(1-p_2)]}\frac{\left(\lambda p_1(1-p_2)\right)^{r_1}}{r_1!}\frac{\left(\lambda p_2(1-p_1)\right)^{r_2}}{r_2!}\cdot \sum_{l=0}^{min{(r_1,r_2)}} \binom{r_1}{l}\binom{r_2}{l}l!\left(\lambda(1-p_1)(1-p_2)\right)^{-l}$$
so that
$$LL(\lambda,p_1,p_2)=-\lambda [1-(1-p_1)(1-p_2)]+\bar{r}_1\cdot \ln(\lambda p_1(1-p_2))+\bar{r}_2\cdot \ln(\lambda p_2(1-p_1))$$
$$-\frac{1}{M}\sum_{m=1}^M \ln(r_{m,1}!r_{m,2}!) 
+\Psi((1-p_1)(1-p_2)\lambda),$$
where
$$\bar{r}_i=\frac{1}{M}\sum_{m=1}^M r_{m,i},\;\;\;i=1,2,$$
$$\Psi(z)=\frac{1}{M}\sum_{m=1}^M \ln \left[\sum_{l=0}^{\min(r_{m,1},r_{m,2})} \binom{r_{m,2}}{l}\binom{r_{m,2}}{l}l!\cdot z^{-l} \right].$$
The maximum likelihood estimators $(\lambda^*,p_1^*,p_2^*)$ are
obtained by maximizing the function $LL$. The first-order conditions for the maximum are
\begin{eqnarray}\label{lle1} \frac{\partial LL}{\partial \lambda}(\lambda,p_1,p_2)
&=&- [1-(1-p_1)(1-p_2)]+\frac{1}{\lambda}\cdot \bar{r}_1 +\frac{1}{\lambda}\cdot \bar{r}_2\\
&+&(1-p_1)(1-p_2)\cdot \Psi'((1-p_1)(1-p_2)\lambda ) =0,\nonumber\\
\label{lle2} \frac{\partial LL}{\partial p_1}(\lambda,p_1,p_2)
&=&-\lambda (1-p_2)+\frac{1}{p_1}\cdot \bar{r}_1-\frac{1}{1-p_1}\cdot \bar{r}_2\\ &-&(1-p_2)\lambda\cdot \Psi'((1-p_1)(1-p_2)\lambda )=0,\nonumber\\
\label{lle3}\frac{\partial LL}{\partial p_2}(\lambda,p_1,p_2)
&=&-\lambda (1-p_1)-\frac{1}{1-p_2}\cdot \bar{r}_1+\frac{1}{p_2}\cdot \bar{r}_2\\&-&(1-p_1)\lambda \cdot \Psi'((1-p_1)(1-p_2)\lambda )=0,\nonumber
\end{eqnarray}
where
$$\Psi'(z)=-\frac{1}{M}\sum_{m=1}^M \frac{\sum_{l=1}^{\min(r_{m,1},r_{m,2})} \binom{r_{m,1}}{l}\binom{r_{m,2}}{l}l!\cdot l\cdot z^{-l-1}}{\sum_{l=0}^{\min(r_{m,1},r_{m,2})} \binom{r_{m,1}}{l}\binom{r_{m,2}}{l}l!\cdot z^{-l}}.$$
Combining (\ref{lle1}) with (\ref{lle2}), and (\ref{lle1}) with (\ref{lle3}) yields
\begin{equation}\label{p12}p_1=\frac{\bar{r}_1}{\lambda},\;\;\;p_2=\frac{\bar{r}_2}{\lambda},\end{equation}
and substituting these equalities into (\ref{lle1}) leads to 
%\begin{equation}\label{eql}\Omega\left(\left(1-\frac{\bar{r}_1}{\lambda}\right)\left(1-\frac{\bar{r}_2}{\lambda}\right)\lambda \right)= \frac{\bar{r}_1\bar{r}_2}{\lambda},\end{equation}
\begin{equation}\label{eql}-\left(\frac{\lambda}{\bar{r}_1}-1\right)\left(\frac{\lambda}{\bar{r}_2}-1\right) \Psi'\left(\left(1-\frac{\bar{r}_1}{\lambda}\right)\left(1-\frac{\bar{r}_2}{\lambda}\right)\lambda \right)= 1,\end{equation}
%where
%$$\Omega(z)=-z\Psi'(z)=\frac{1}{M}\sum_{m=1}^M \frac{\sum_{l=1}^{\min(r_{m,1},r_{m,2})} l!\binom{r_{m,1}}{l}\binom{r_{m,2}}{l}\cdot l\cdot z^{-l}}{\sum_{l=0}^{\min(r_{m,1},r_{m,2})} l!\binom{r_{m,1}}{l}\binom{r_{m,2}}{l}z^{-l}}.$$
We thus have
\begin{theorem}
	The maximum likelihood estimators $(\lambda^*,p_1^*,p_2^*)$ are obtained by solving (\ref{eql}) for $\lambda^*$,
	and using (\ref{p12}).
	
\end{theorem}
Equation (\ref{eql}) can be solved numerically, that is by an iterative method. We note that as a starting point for the iterations,
one can use the estimator $\hat{\lambda}$ obtained by the moment method.

It should be noted that in order for the solution given above to be in the domain $\lambda>0$, $p_1,p_2\in [0,1]$, it is necessary, by
(\ref{p12}), that $\lambda^*>\max(\bar{r}_1,\bar{r}_2)$. In our numerical simulations we have observed that in some cases the equation (\ref{eql}) does {\it{not}} have a solution - corresponding to the fact that the likelihood function attains its maximum at a boundary point
of the relevant domain $\lambda\geq 0$, $p_1,p_2\in [0,1]$, {\it{e.g.}} with $\hat{p}_i=1$. This pathology occur when the true value of 
$p_1$ or of $p_2$ is close to the boundary, and $M$ is relatively small. However as the amount of data $M$ increases the probability of such an occurance goes to $0$, indeed from the general theory of maximum likelihood estimation it follows that as $M\rightarrow \infty$,
maximum likelihood estimator close to the true values of the parameters will exist with probability approaching $1$ (see Theorem 
\ref{norm} below). 

A related issue is the {\it{uniqueness}} of the solution of (\ref{eql}), assuming it exists. In all of our experiments we have found that, whenever a solution
$\lambda^*>\max(\bar{r}_1,\bar{r}_2)$ of (\ref{eql}) exists, it is unique. However we have not been able to prove this 
fact analytically, and finding such a proof remains an open question.

\begin{table}\label{MLvar}
	\begin{center}
		\begin{tabular}{|c|c|c|}
			%	\hline {} & {$\hat{\lambda}$} & {$\hat{p}_1$} & {$\hat{p}_2$} \\ 
			\hline {M}  & {$Std(\lambda^*)$ (sim.)} & {$Std(\lambda^*)$ (app.)} \\ 
			\hline	100  & 1.33 & 1.22\\
			\hline	200 & 0.90 & 0.86 \\
			\hline	500  & 0.56 & 0.55\\
			\hline 
		\end{tabular}
	\end{center} 
	\caption{Standard deviation of the maximum likelihood estimator $\lambda^*$ based on $5000$ simulations, and the 
		corresponding approximations, for parameters $\lambda=10$, $p_1=0.4$, $p_2=0.7$.}
\end{table}

\subsection{Asymptotic normality and the variance of the maximum-likelihood estimators}

The general theory of the asymptotic behavior of maximum-likelihood estimators entails the convergence of 
the distribution of our estimators to a normal one. Theorem 5.1 in Ch. 6 of Lehman \& Casella \cite{lehman}, whose hypotheses are
easily verified for our problem, implies:
\begin{theorem}\label{norm}
	With probability tending to $1$ as $M\rightarrow \infty$, there exist
	ML estimators $(\lambda^*,p_1^*,p_2^*)$, with 
	$$\sqrt{M}(\lambda^*-\lambda)\xrightarrow{{\cal{L}}} N(0,[I^{-1}]_{11}),$$
	$$\sqrt{M}(p_1^*-p_1)\xrightarrow{{\cal{L}}} N(0,[I^{-1}]_{22}),\;\;\sqrt{M}(p_2^*-p_2)\xrightarrow{{\cal{L}}} N(0,[I^{-1}]_{33})$$
	as $M\rightarrow \infty$, where the arrows denote convergence in distribution, and $I$ is Fisher's information matrix.
\end{theorem}

The Fisher information matrix is given by
 \begin{small}
 $$I=I(\lambda,p_1,p_2)=-\left(\begin{array}{ccc}
E\left(\frac{\partial^2}{\partial \lambda^2 }\ln P(r_1,r_2) \right) & E\left(\frac{\partial^2}{\partial \lambda \partial p_1 }\ln P(r_1,r_2)\right) &  E\left(\frac{\partial^2}{\partial \lambda \partial p_2 }\ln P(r_1,r_2)\right)\\ 
\cdot  & E\left(\frac{\partial^2}{ \partial p_1^2 }\ln P(r_1,r_2)\right) &  E\left(\frac{\partial^2}{\partial p_1 \partial p_2 }\ln P(r_1,r_2)\right)\\ 
\cdot & \cdot &  E\left(\frac{\partial^2}{ \partial p_2^2 }\ln P(r_1,r_2)\right)
 \end{array}  \right)$$
 \end{small}
(sub-diagonal elements determined by symmetry), where in the expectations it is assumed that $(r_1,r_2)$ are distributed according to 
the model with parameters $(\lambda,p_1,p_2)$, hence, for example
$$I_{11}=-E\left(\frac{\partial^2}{\partial \lambda^2 }\ln P(r_1,r_2) \right)= \sum_{r_1=0}^\infty \sum_{r_2=0}^\infty P(r_1,r_1)\frac{\partial^2}{\partial \lambda^2 }\ln P(r_1,r_2).$$
In the Supporting Material (Gadrich and Katriel \cite{gadrich}, Section 3) we give expressions for evaluating the elements of the matrix $I$. These 
expressions involve infinite sums which, in actual computation, must be approximated by truncating to a finite sum.
We note that expressions for the elements of the Fisher matrix of the Bivariate Poisson distribution with respect to the  
parameters $\theta_1,\theta_2,\theta_{12}$ (see section \ref{model}) were obtained by Holgate \cite{holgate} (see also Kocherlakota \& Kocherlakota \cite{kocherlakota}), though we did not use these in our derivations.

In particular, we have asymptotic expressions for the variances of the ML estimators
\begin{eqnarray}\label{vml}VAR(p_1^*)&=&[I^{-1}]_{22}\cdot \frac{1}{M}+O\left(\frac{1}{M^2} \right),\;\;VAR(p_2^*)=[I^{-1}]_{33}\cdot \frac{1}{M}+O\left(\frac{1}{M^2} \right),\nonumber\\
VAR(\lambda^*)&=&[I^{-1}]_{11}\cdot \frac{1}{M}+O\left(\frac{1}{M^2} \right).
\end{eqnarray}
Table 2 shows the approximate variances of the ML estimator for $\lambda$ (for $\lambda=10,p_1=0.7,p_2=0.4$) obtained using (\ref{vml}), as well as the variance obtained from $5000$ simulations, for several values of $M$. It can be observed that as $M$
grows larger, the asymptotic approximation agrees with the simualtion results. 

Using the above expressions for the variances of the estimators, approximate confidence intervals for the ML estimators can
be computed in the standard way.

\subsection{Comparing maximum likelihood and moment estimators}

\begin{figure}
	\centering
	\includegraphics[width=0.48\linewidth]{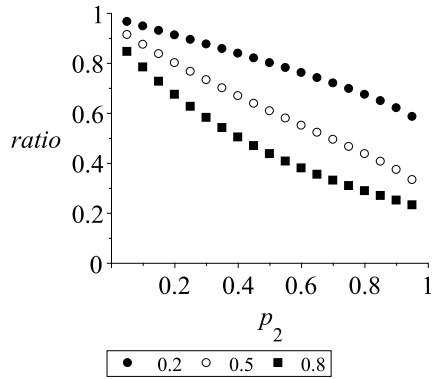}	
	\caption{The asymptotic ($M\rightarrow \infty$) ratio of the standard deviation of the maximum likelihood estimator to that of the moment estimator for $\lambda$, for different values of $p_1,p_2$, fixing $\lambda=10$. Values for $p_1$ were 
	$0.2$ (solid circles), $0.5$ (empty circles), $0.8$ (squares), and for each value of $p_1$, $p_2$ ranges from $0.05$ to $0.95$.}
	\label{compp}
\end{figure}

The analytic asymptotic expressions for the variances of the moment estimators (Theorem \ref{variance}) and 
for the maximum likelihood estimators (\ref{vml}) allow us to perform a comparison of the accuracy of the two estimators,
in the limit in which the number of observations $M$ is large. Note that the standard deviation of both estimators decreases 
like $\frac{1}{\sqrt{M}}$, so that the comparison involves the corresponding pre-factor. We concentrate on the 
estimator for $\lambda$, which is of prime interest.  Figure \ref{compp} shows the 
results of such a comparison.  We fixed $\lambda=10$ and plotted the ratio of the standard deviation of the 
ML estimator to that of the moment estimator for different values of $p_1,p_2$. We see that the maximum
likelihood estimator is always more accurate than the moment estimator, and the effect is more pronounced as $p_1,p_2$ increase.
Thus, for example, when $p_1=0.5$ and $p_2=0.2$, the ratio of the standard deviations is $0.80$, while if $p_1=0.5$ and $p_2$ is
increased to $0.9$ the maximum likelihood estimator is nearly $3$ times as accurate as the moment estimator. 

Considering the effect of the value of $\lambda$ on the ratio of the standard deviations of the ML and moment estimates,
we found that this dependence is very weak, {\it{e.g.}}, taking $p_1=0.7,p_2=0.4$,
the ratio takes the value $0.5605$ for $\lambda=10$ and $0.5621$ for $\lambda=100$.

To verify these observations in the non-asymptotic case, now use a simulation test to compare the accuracy of the two estimation methods we have developed. 
In our experiments, we generated data on the number of defects detected by each inspector in $M$ items ($M=100,200,500$), using the model, with defect rate $\lambda=10$, and with detection rates $p_1=0.4$ and $p_2=0.7$. Using the simulated data, we computed the estimators $(\hat{\lambda},\hat{p}_1,\hat{p}_2)$ using the moment methods, and the estimators $(\lambda^*,p_1^*,p_2^*)$ using the
maximum likelihood method. We repeated this $5000$ times.

\begin{figure}
	\centering
	\includegraphics[height=5cm,width=0.48\linewidth]{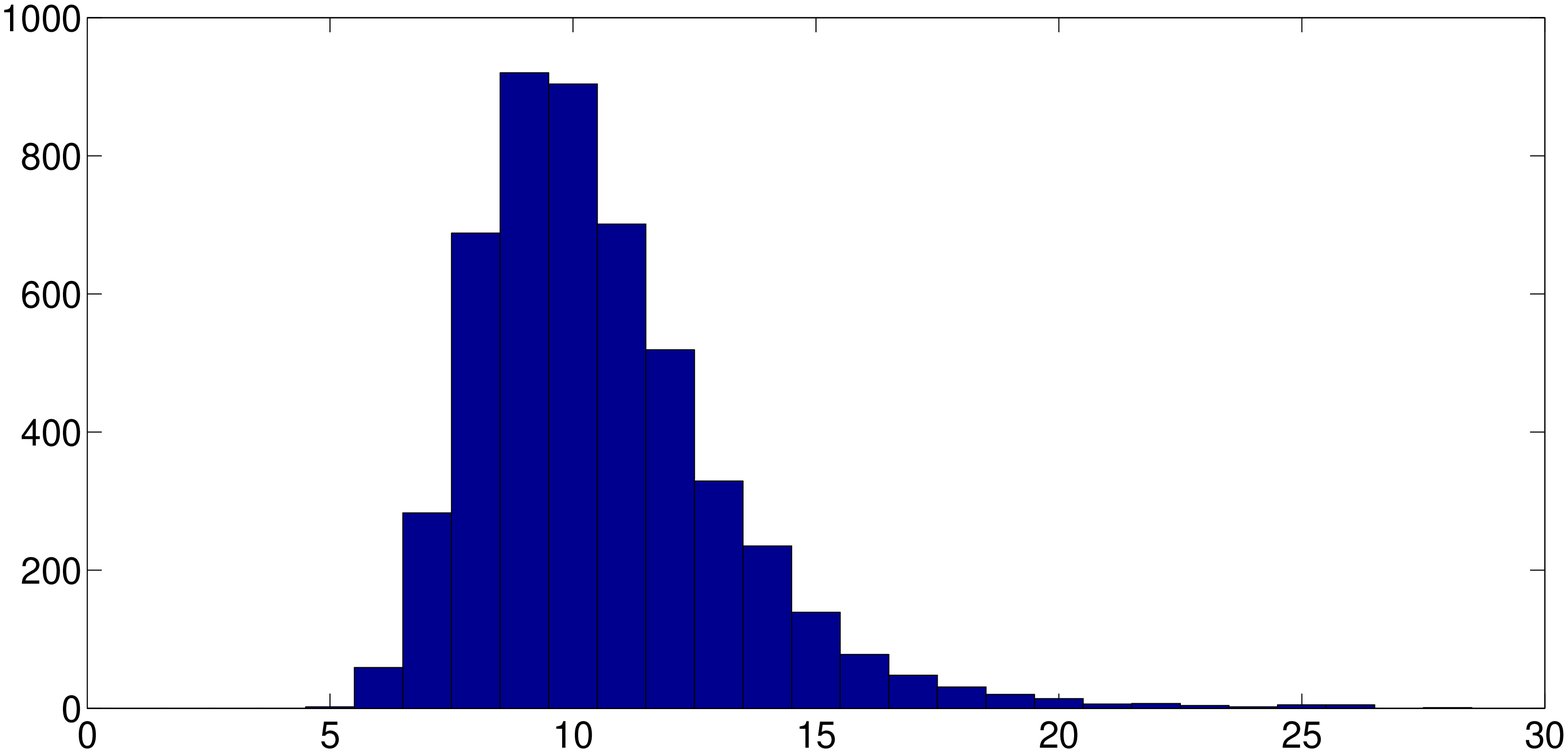}
	\includegraphics[height=5cm,width=0.48\linewidth]{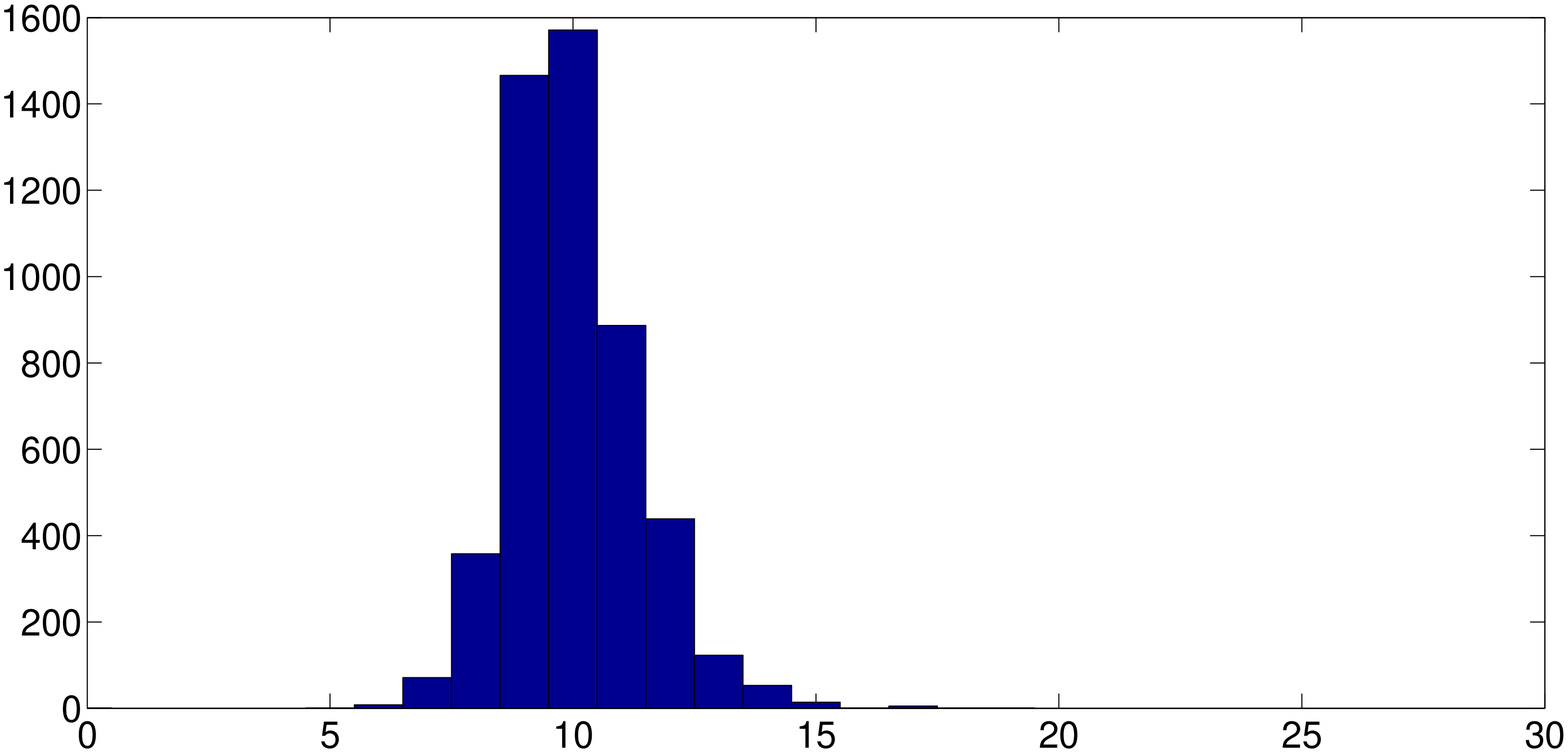}\\
	\includegraphics[height=5cm,width=0.48\linewidth]{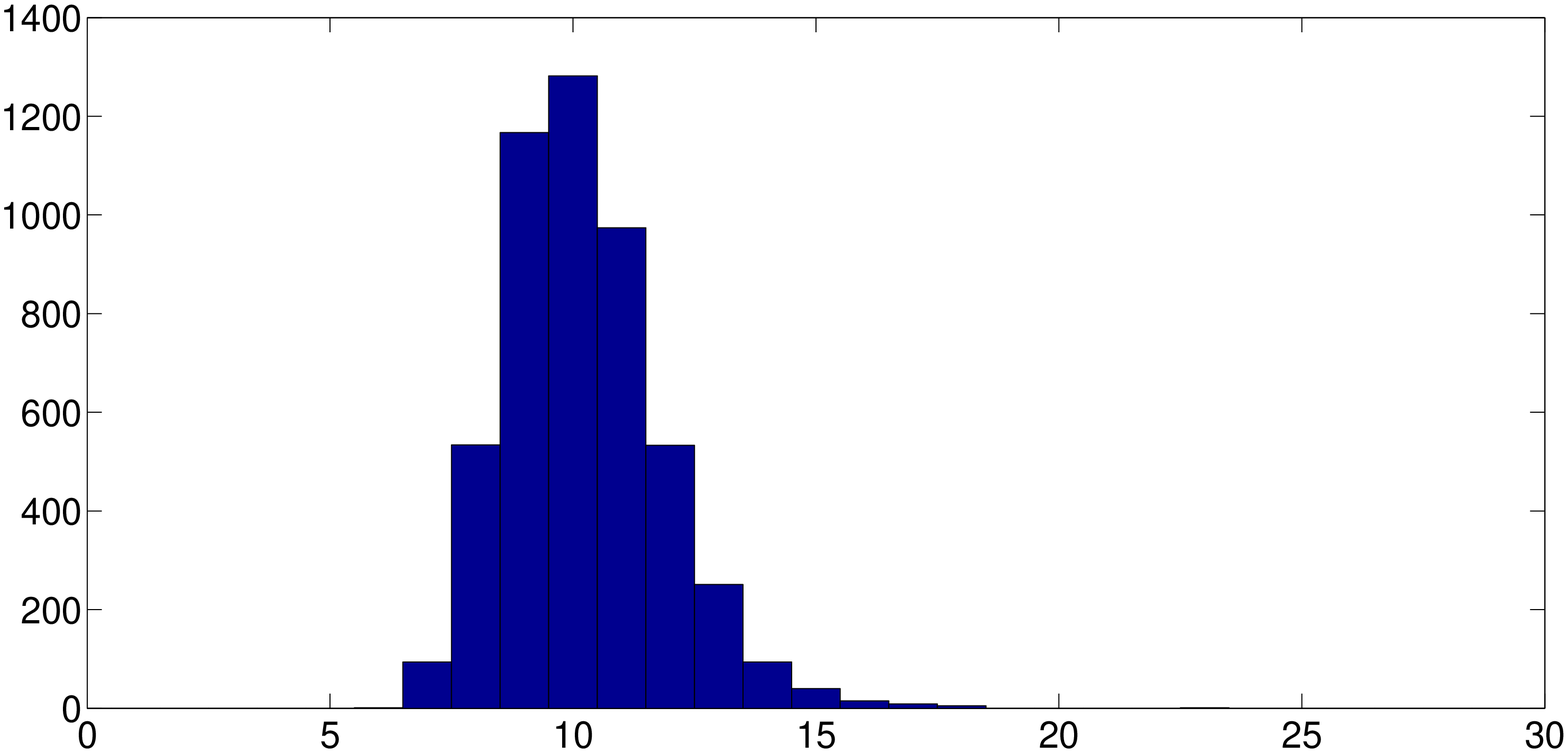}
	\includegraphics[height=5cm,width=0.48\linewidth]{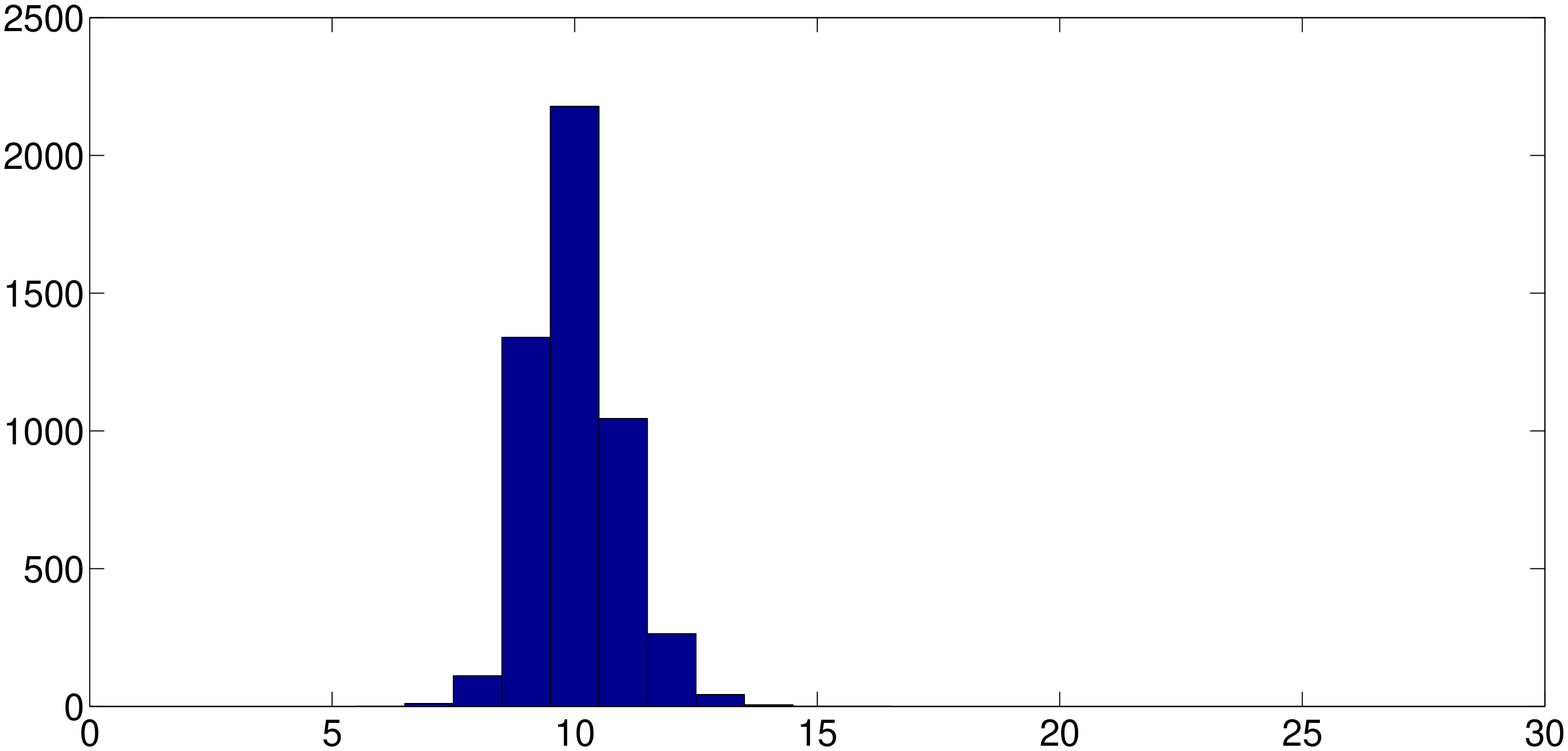}\\
	\includegraphics[height=5cm,width=0.48\linewidth]{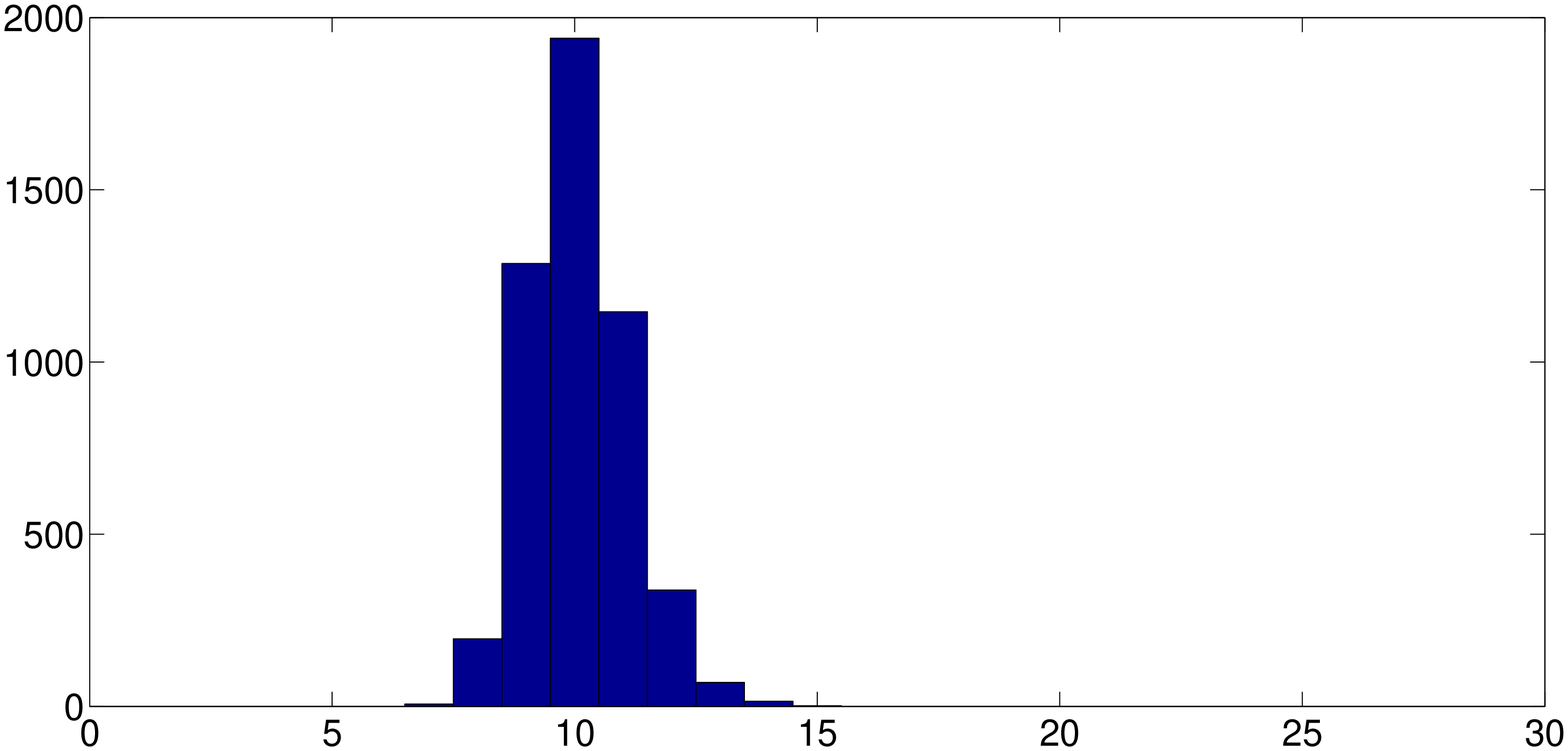}
	\includegraphics[height=5cm,width=0.48\linewidth]{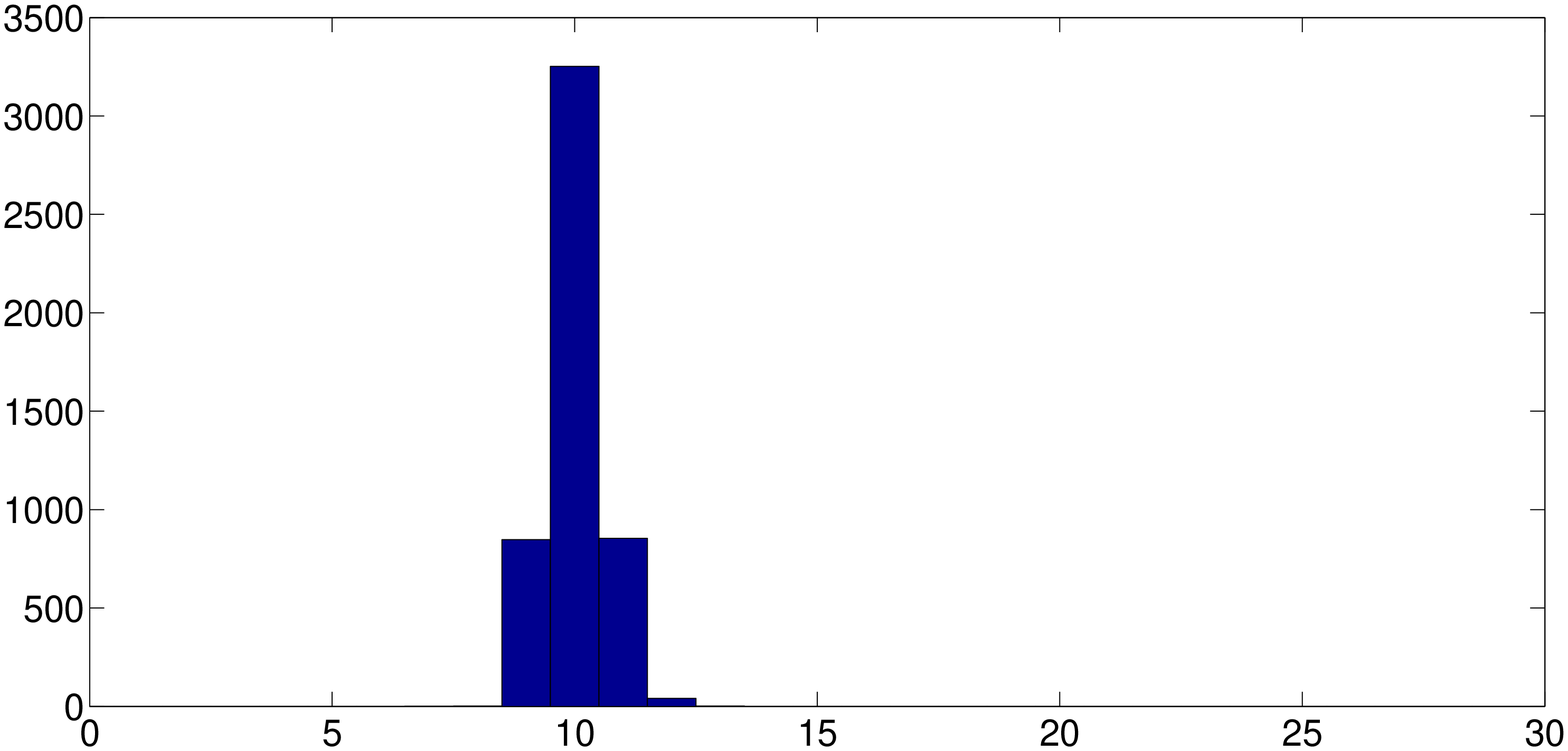}	
	\caption{Histograms of estimates of $\lambda$ using the moment method (left) and the maximum likelihood method (right),
		for $M=100$ (top), $M=200$ (middle), $M=500$ (bottom).
		In all
		cases, $5000$ tests were performed with $\lambda=10,p_1=0.4,p_2=0.7$.}
	\label{fig2}
\end{figure}

\begin{table}[]
%	\centering
	\label{tab2}
	\begin{center}
	\begin{tabular}{|l|l|l|l|l|l|}
		\hline
		M   & \multicolumn{2}{c|}{$\lambda$ mean} & \multicolumn{2}{c|}{\begin{tabular}[c]{@{}c@{}}$\lambda$\\ standard deviation\end{tabular}} & \multicolumn{1}{c|}{\multirow{2}{*}{\begin{tabular}[c]{@{}c@{}}$\%$ of trials where \\ ML is better \end{tabular}}} \\ \cline{1-5}
		& moment   & \multicolumn{1}{c|}{ML}  & moment                               & \multicolumn{1}{c|}{ML}                              & \multicolumn{1}{c|}{}                                                                           \\ \hline
		100 & $10.54$  & $10.00$                  & $2.61$                               & $1.33$                                               & $71.5$                                                                                          \\ \hline
		200 & $10.21$  & $10.02$                  & $1.60$                               & $0.90$                                               & $68.8$                                                                                          \\ \hline
		500 & $10.08$  & $10.01$                  & $1.01$                               & $0.56$                                               & $70.3$                                                                                          \\ \hline
	\end{tabular}
\end{center}
	\caption{Results of simulation tests of the moment estimator and the maximum likelihood (ML) estimator. For each of the three rows,  $N=5000$ data sets were
		generated with parameter values $\lambda=10,p_1=0.4,p_2=0.7$. }
\end{table}

Figure \ref{fig2} displays the histograms of the estimates obtained for $\lambda$ using each of the two methods.
As can be seen in the figure, the estimates made using the maximum likelihood method are on the average more precise.

Characteristics of the distributions of the estimators are presented in Table 3. 
We see that the maximum likelihood estimator is superior both in the sense that its bias is smaller and in that
its variance is smaller. As can be seen in the right-most column, for approximately $70\%$ of the $5000$ data samples
the estimate obtained by the maximum likelihood estimator was closer to the true value than that obtained by the
moment estimator.

We thus conclude that for our problem, despite the fact that the moment estimator is very `natural' and 
simple, the maximum likelihood estimator, which is harder to compute, performs better.

\section{Comparison with Capture-Recapture estimation}
\label{comparison}

The methods developed in this work are aimed at estimating defect and detection rates when 
no information is available regarding the number of defects which were jointly detected.
When such information {\it{is}} available, one can use the 
Capture-Recapture method (Bonett \cite{bonnet}, Briand et al. \cite{briand}, Petersson et al. \cite{petersson}). This method originated as method of estimating the size of animal populations
by capturing some animals, marking and releasing them, and then capturing another group of animals and counting 
how many of these are marked (Chao \cite{chao2}, McCrea and Morgan {\cite{mcrea}). 

It is intuitively clear that using the additional information on joint detections should result
in more precise estimates, and we would now like to quantify this intuition by comparing the variance 
of our estimators with that of the estimator obtained from the Capture-Recapture method.
%In the case of our moment-type estimator, we can make the comparison analytically, using the expressions for its
%variance found in Section \ref{moment}, while in the case of the maximum-likelihood estimators, for which we do not have 
%explicit expressions for the variance, the comparison can be performed by simulation.

The prerequisite for applying the Capture-Recapture method is availability of the values of the variables $X_{m,1},X_{m,2},Y_m$:
the number of defects detected only by inspector $1$, only by inspector $2$, and by both inspectors, respectively.
These are independent and distributed according to (\ref{x}),(\ref{y}).

Estimators for $p_1,p_2,\lambda$ are then given by (see Gadrich \& Katriel \cite{gadrich}, Section 4)
$$\hat{p}_{1,CR}=\frac{\bar{y}}{\bar{x}_2+\bar{y}},\;\;\;\hat{p}_{2,CR}=\frac{\bar{y}}{\bar{x}_1+\bar{y}},\;\;\hat{\lambda}_{CR}=\frac{(\bar{x}_1+\bar{y})(\bar{x}_2+\bar{y})}{\bar{y}},$$
where
$$\bar{x}_i=\frac{1}{M}\sum_{m=1}^M x_{m,i},\;\;i=1,2,\;\;\;\;\bar{y}=\frac{1}{M}\sum_{m=1}^M y_{m}.$$

These estimators can be obtained both using a moment-type approach and using a maximum-likelihood approach - 
in contrast with the problem which is the main focus of our work, here the two approaches yield identical estimators.

Note that here it is of no importance that the defects occur on $M$ separate items:  the data from
all items is summed up, and no loss would be incurred if only the aggregate data on number of defects detected by each of the
inspectors seperately, and by both of them, were available.
 This is different from the estimation problem studied in this paper, in which having data from several items is essential for identifiability. In this connection, it should be noted that while the Capture-Recapture method can be 
 used to assess the quality of {\it{individual}} items, as is in fact done in the case of software testing where a 
 particular program is being inspected. Our method, since it requires $M\geq 2$ items, is appropriate for estimating the
 {\it{rate}} of defects $\lambda$ and not the quality of individual items, and is thus suited for acceptance sampling rather
 than the testing of individual items.

We analyzed the bias and variance of the above estimators - although the Capture-Recapture method is well known, we were
not able to locate the results given below in the literature, since our framework, in which we are estimating a defect rate, is different from the common framework in which a fixed number of total defects is being estimated.

We have (see Gadrich and Katriel \cite{gadrich}, Section 4, for the calculations)
\begin{theorem}\label{expmse}
	$$E(\hat{p}_{1,CR})=p_1,\;\;\; E(\hat{p}_{2,CR})=p_2,$$
	that is the estimators $\hat{p}_{1,CR},\hat{p}_{2,CR} $ are unbiased.
	
	For $\hat{\lambda}_{CR}$ we have, as $M\rightarrow \infty$,
	$$E(\hat{\lambda}_{CR})=\lambda + 
	\left(\frac{1}{p_1}-1 \right)\left(\frac{1}{p_2}-1 \right)
	\cdot \frac{1}{M} +O\left(\frac{1}{M^2} \right).$$
\end{theorem}

\begin{theorem}\label{variancemse}
	As $M\rightarrow \infty$,
	\begin{eqnarray}\label{varms1}
	VAR(\hat{p}_{i,CR})&=& \frac{p_i^2(1-p_i)}{ \lambda p_1 p_2}\cdot \frac{1}{M}+O\left(\frac{1}{M^2}\right),\;\;i=1,2,\;\;
	\end{eqnarray}
	\begin{equation}\label{varms3}
	VAR(\hat{\lambda}_{CR})=\lambda\cdot \left(1+\left(\frac{1}{p_1}-1 \right)\left(\frac{1}{p_2}-1 \right)\right)\cdot \frac{1}{M}+O\left(\frac{1}{M^2}\right).
	\end{equation}
\end{theorem}

Combining Theorems \ref{exp} and \ref{expmse} yields
$$E(\hat{\lambda})-E(\hat{\lambda}_{CR})=\frac{ \lambda (p_1p_2+1)  }{p_1 p_2}\cdot \frac{1}{M}+O\left(\frac{1}{M^2}\right),$$
showing that the bias of our moment estimator is greater than the of the Capture-Recapture estimator.
Combining Theorems \ref{variance} and \ref{variancemse} yields
%$$VAR(\hat{p}_1)-VAR(\hat{p}_{1,MS})=\frac{p_1}{ p_2}  \left(  p_1p_2+ 1 \right) \cdot \frac{1}{M} + O\left(\frac{1}{M^2}\right),$$
%$$VAR(\hat{p}_2)-VAR(\hat{p}_{2,MS})=\frac{p_2}{ p_1}  \left(  p_1p_2+ 1 \right) \cdot \frac{1}{M} + O\left(\frac{1}{M^2}\right),$$
$$VAR(\hat{\lambda})-VAR(\hat{\lambda}_{CR})= \frac{\lambda^2(p_1p_2+1)}{p_1p_2}\cdot \frac{1}{M}+O\left(\frac{1}{M^2}\right),$$
showing that, as expected, the accuracy of Capture-Recapture estimations is better,
when it is applicable. It can be checked that similar results hold for the bias and variance of the estimators for $p_1,p_2$.

As a numerical example, if we take $\lambda=10,p_1=0.4,p_2=0.7,M=200$, then the bias of the Capture-Recapture estimator for $\lambda$ 
is $0.003$, while that of our moment estimator is $0.25$, and of our maximum likelihood estimator (based on simulation)
is $0.02$.
The standard deviation of the Capture-Recapture estimator for $\lambda$ is
$0.2$, whereas our moment estimator gives a standard deviation of $1.6$ and our maximum-likelihood estimator gives a standard 
deviation (using (\ref{vml}) and verified by simulation) of $0.9$. We conclude that the Capture-Recapture is considerably more precise than our estimators for a given sample size of items, 
which as noted above, is not surprising in view of the fact that it uses more information.

Thus the estimators we develop should be useful when the information required to apply the Capture-Recapture method is {\it{not}} available,
that is when we do not have counts of the number of joint detections.

Two necessary assumptions for applying the model proposed here should be kept in mind:
\begin{itemize}
	\item [(1)] The number of defects per item in the population from which the items are sampled is Poisson-distributed with a fixed parameter $\lambda$. This requires that the manufacturing process by which the items were produced is stable in the sense that there are no systematic changes in the process over the time period during which the items were produced.
	\item [(2)] The records made by the two inspectors are independent in the sense that the fact 
	that a defect is recorded by one of the observers does not change the probability that the same event is recorded
	by the other inspector. In particular this assumption would not hold if defects are heterogeneous in the sense that some defects are 
	easier to detect than others, which would mean that a defect that has been detected by one inspector is more likely to be
	detected by the other. 
\end{itemize}

We note that these assumptions also underlie the Capture-Recapture method when applied to two inspectors.

%\section*{Acknowledgement}
%We are grateful to the referees for their valuable comments and suggestions.


\begin{thebibliography}{3}
	% Format for journal reference
	% \bibitem[Author (1999)]{RefJ}
	% Author, I. (1999). Paper Title. Journal Title, Vol, pp--pp
	%
	
	%
	% Format for book reference
	% \bibitem[Author and Smith(2001)]{RefB}
	% Author, I., \& Smith, J. K. (2001) Chapter title. In Name (Ed.)
	% Book title (pp. xx--yy). Place, Publisher
	%
	% etc
	
%	\bibitem{bezerra}
%	Bezerra, Kátia F., et al. (2011).  The number of street children and adolescents in two cities of Brazil using Capture-Recapture. 
%	{\em Journal of paediatrics and child health}, 47,  524--529.‏	
%	
%	\bibitem{bishop} Bishop, Y.M., Fienberg, S.E.  and Holland, P.W. (2007). {\em Discrete multivariate analysis: theory and practice }. New-York, Springer Science \& Business Media.
	
%	\bibitem{boden}
%	Boden, L.I. and Ozonoff, A.L. (2008) Capture-recapture estimates of nonfatal workplace injuries and illnesses. {\em Annals of epidemiology}, 18, 500--506.
%	
	\bibitem{bonnet}
	Bonett, D. G. (1988). Estimating the number of defects under imperfect inspection. 
	{\em Journal of Applied Statistics} 15: 63--67.
	
	\bibitem{bonnet1}
	Bonett, D. G. \&  Woodward, J. A. (1994). Sequential defect removal sampling. {\em Management Science} 40: 898-902.‏
	
	\bibitem{briand}
	Briand, L. C., El Emam, K., Freimut, B. G., \& Laitenberger, O. (2000). A comprehensive evaluation of capture-recapture models for estimating software defect content. {\em IEEE Transactions on Software Engineering} 26: 518--540.
	
%	\bibitem{chang}
%	Chang, Y.F. et al. (1997).  Bite incidence in the city of Pittsburgh: a capture-recapture approach. {\em American Journal of Public %Health},  87, 1703--1705.‏
	
	
%	\bibitem{chao1} Chao, A., et al. (2001). The applications of capture-recapture models to epidemiological data. 
%	{\em Statistics in Medicine}, 20, 3123--3157 .‏
%	
	\bibitem{chao2}
	Chao, A. (2001) An overview of closed capture-recapture models. {\em Journal of Agricultural, Biological, and Environmental Statistics} 6:  158--175.‏
	
	\bibitem{chun} Chun, Y. H. (2005). Serial inspection plan in the presence of inspection errors: Maximum likelihood and maximum entropy approaches. {\em Quality Engineering} 17: 627-632.‏
%	\bibitem{fisher}
%	Fisher, N., et al. (1994) Estimated numbers of homeless and homeless mentally ill people in north east Westminster by using capture-recapture analysis. {\em BMJ}, 308,  27--30.‏
%	
%	\bibitem{gill}
%	Gill, G. V., Ismail, A.A. and Beeching, N.J. (2001).  Use of capture-recapture techniques in determining the prevalence of type 2 %diabetes. {\em QJM}, 94,  341--346.‏
	‏
%	\bibitem{goldman}
%	Goldman, G.S. (2003). Using capture-recapture methods to assess varicella incidence in a community under active surveillance.
%	{\em Vaccine}, 21, 4250--4255 .
%	
\bibitem{gadrich} Gadrich, T. \& Katriel, G. (2018). Supporting Material for the paper ``Estimating the rate of defects under imperfect sampling inspection - a new approach",\\
{\small{{\url{https://drive.google.com/file/d/14Kku7cOUgKuZjqJebHwfBhANff_ommr3/view?usp=sharing}}}}.

\bibitem{grubbs} Grubbs, F. E. (1948). On estimating precision of measuring instruments and product variability. {\em{Journal of the American Statistical Association}} 43: 243--264.‏

\bibitem{grubbs1} Grubbs, F. E. (1973). Errors of measurement, precision, accuracy and the statistical comparison of measuring instruments. {\em{Technometrics}}, 15: 53--66.

	\bibitem{holgate} Holgate, P. (1964). Estimation for the bivariate Poisson distribution. {\em Biometrika}  51: 241--287.
	
	\bibitem{johnson} Johnson, N. L., Kotz, S. \&  Balakrishnan, N. (1997). {\em Discrete multivariate distributions}.  New-York: Wiley.
	
%	\bibitem{joun}
%	Jouanjus, E., et al. (2012). Use of multiple sources and capture-recapture method to estimate the frequency of hospitalizations related %to drug abuse. {\em Pharmacoepidemiology and drug safety}, 21, 733--741.
	
%	\bibitem{khabsa}
%	Khabsa, M., Giles, C.M. (2014). The number of scholarly documents on the public web. {\em PloS One}, 9.5, e93949.
%	

\bibitem{kocherlakota} Kocherlakota, S. \& Kocherlakota, K. (2004). {\em{Bivariate discrete distributions}}. Encyclopedia of Statistical Sciences 1, New-York: Marcel Dekker.

	\bibitem{kotz}
	Kotz, S. \&  Johnson, N. L. (1984). Effects of false and incomplete identification of defective items on the reliability of acceptance sampling. {\em Operations Research} 32: 575--583.
	
	\bibitem{lehman} Lehmann, E. L. \& Casella, G. (2006). {\em{Theory of point estimation}}. New-York: Springer.‏
%	\bibitem{laporte}
%	LaPorte, R.E. (1994). Assessing the human condition: capture-recapture techniques. {\em BMJ}, 308, 5.‏
%	
%	\bibitem{lum}
%	Lum, K., Peice, M.E., and Banks, D. (2013). Applications of multiple systems estimation in human rights research. 
%	{\em The American Statistician}, 67, 191--200.
%	
	%\bibitem{manrique}
	%Manrique-Vallier, Daniel, Megan E. Price, and Anita Gohdes. "Multiple systems estimation techniques for estimating casualties in armed %conflicts." In: Seybolt, Taylor B., Jay D. Aronson, and Baruch Fischhoff, eds., Counting Civilian Casualties: An Introduction to Recording %and Estimating Nonmilitary Deaths in Conflict. Oxford University Press, 2013.‏
	
%	\bibitem{mastro}
%	Mastro, T.D. et al. (1994).  Estimating the number of HIV-infected injection drug users in Bangkok: a capture--recapture method. 
%	{\em American Journal of Public Health}, 84,  1094--1099.‏		
%	

\bibitem{lombard}
Lombard, F., \& Potgieter, C. J. (2012). Another look at the Grubbs estimators. 
{\em{Chemometrics and Intelligent Laboratory Systems}}. 110: 74--80.‏

	\bibitem{mcrea}
	McCrea, R.S. \& Morgan, B.J.T. (2014) {\em Analysis of capture-recapture data}. Boca Raton:  CRC Press.
	
	\bibitem{montgomery}
	Montgomery, D. C. (2009). {\em Statistical quality control}. New York: Wiley.‏
	
	\bibitem{mood} 
	Mood, G., Graybill, F. \& Boes, D.C. (1974). {\em Introduction to the Theory of Statistics}, New-York: McGraw-Hill.
	
%	\bibitem{oosterlie}
%	Oosterlee, A., Vink, R.M, and Smit, F. (2009). Prevalence of family violence in adults and children: estimates using the capture-recapture method. {\em The European Journal of Public Health}, 19,  586--591.
%	
	\bibitem{petersson}
	Petersson, H., Thelin, T., Runeson, P.,\& Wohlin, C. (2004). Capture-recapture in software inspections after 10 years research - theory, evaluation and application. {\em Journal of Systems and Software} 72: 249--264.‏
	
%	\bibitem{razzak}
%	Razzak, J.A., Luby, S.P. (1998). Estimating deaths and injuries due to road traffic accidents in Karachi, Pakistan, through the capture-%recapture method. {\em International Journal of Epidemiology} 27, 866--870.‏
	
%	\bibitem{trader}
%	Trader, R. L.,  Fenwick Huss, H. (1985). Prediction in the presence of imperfect inspections. {\em Communications in Statistics-%Simulation and Computation}, 14(2), 425-440.‏
%	
\end{thebibliography}
\end{document}